\documentclass[prd,twocolumn,superscriptaddress,floatfix, preprintnumbers,nofootinbib]{revtex4-2} 
\usepackage{epsfig}
\usepackage{graphicx}
\usepackage{hyphenat}
\usepackage{amsmath}
\usepackage{amssymb}
\usepackage{mathtools}
\usepackage{mathrsfs}
\usepackage{slashed}
\usepackage{epstopdf}
\usepackage{xcolor}
\usepackage{braket}
\usepackage{float}
\definecolor{lcolor}{rgb}{0.,0.0,0.}
\definecolor{citcolor}{rgb}{0,0.,0.5}
\usepackage[breaklinks,colorlinks,urlcolor=blue,citecolor=blue,linkcolor=blue]{hyperref}
\usepackage{multirow}
\usepackage{soul}

\newcommand{\beq}{\begin{eqnarray}}
\newcommand{\eeq}{\end{eqnarray}}

\newcommand{\bem}{\begin{multline}}
\newcommand{\eem}{\end{multline}}
\newcommand{\beg}{\begin{gather}}
\newcommand{\eeg}{\end{gather}}

\newcommand{\nn}{\nonumber\\}

\newcommand{\ben}{\begin{eqnarray*}}
\newcommand{\een}{\end{eqnarray*}}

\newcommand{\secn}[1]{Section~1}
\newcommand{\appn}[1]{Appendix~1}

\long\def\comment#1{ }

\def\and{\quad\text{and}\quad}

\def\0{{\boldsymbol 0}}

\def\0{{\boldsymbol 0}}

\begin{document}

\title{Towards a Real-Time Computation of Timelike Hadronic Vacuum Polarization and Light-by-Light Scattering: Schwinger Model Tests}

\author{João Barata}
\email[]{jlourenco@bnl.gov}
\affiliation{Physics Department, Brookhaven National Laboratory, Upton, NY 11973, USA}
\author{Kazuki Ikeda}
\email[]{kazuki.ikeda@umb.edu}
\affiliation{Department of Physics, University of Massachusetts Boston, Boston, MA 02125, USA}
\affiliation{Co-design Center for Quantum Advantage}
\affiliation{Center for Nuclear Theory, Department of Physics and Astronomy,
Stony Brook University, Stony Brook, New York 11794-3800, USA}
\author{Swagato Mukherjee}
\email[]{swagato@bnl.gov}
\affiliation{Physics Department, Brookhaven National Laboratory, Upton, NY 11973, USA}
\author{Jonathan Raghoonanan}
\email[]{jraghoona@bnl.gov}
\affiliation{Physics Department, Brookhaven National Laboratory, Upton, NY 11973, USA}

\begin{abstract}
Hadronic vacuum polarization (HVP) and light-by-light scattering (HLBL) are crucial for evaluating the Standard Model predictions concerning the muon's anomalous magnetic moment. However, direct first-principle lattice gauge theory-based calculations of these observables in the timelike region remain challenging. Discrepancies persist between lattice quantum chromodynamics (QCD) calculations in the spacelike region and dispersive approaches relying on experimental data parametrization from the timelike region. Here, we introduce a methodology employing 1+1-dimensional quantum electrodynamics (QED), i.e. the Schwinger Model, to investigate the HVP and HLBL. To that end, we use both tensor network techniques, specifically matrix product states, and classical emulators of digital quantum computers. Demonstrating feasibility in a simplified model, our approach sets the stage for future endeavors leveraging digital quantum computers.
\end{abstract}

\maketitle

\section{Introduction}\label{sec:intro}

The determination of the muon's anomalous magnetic moment is one of the most ambitious and important high-precision programs related to the Standard Model (SM) of particle physics~\cite{Aoyama:2020ynm,Jegerlehner:2009ry}. The quantum corrections to the classical (Dirac) term to the muon's magnetic moment couple to all three sectors, QED, Weak, and QCD, of the SM and, therefore, any discrepancies between experiment and theory give a clear signal of beyond the SM physics. The comparisons between the experimental measurements and the theoretical predictions are still being actively studied.

\begin{figure}
    \centering
    \includegraphics[width=0.45\textwidth]{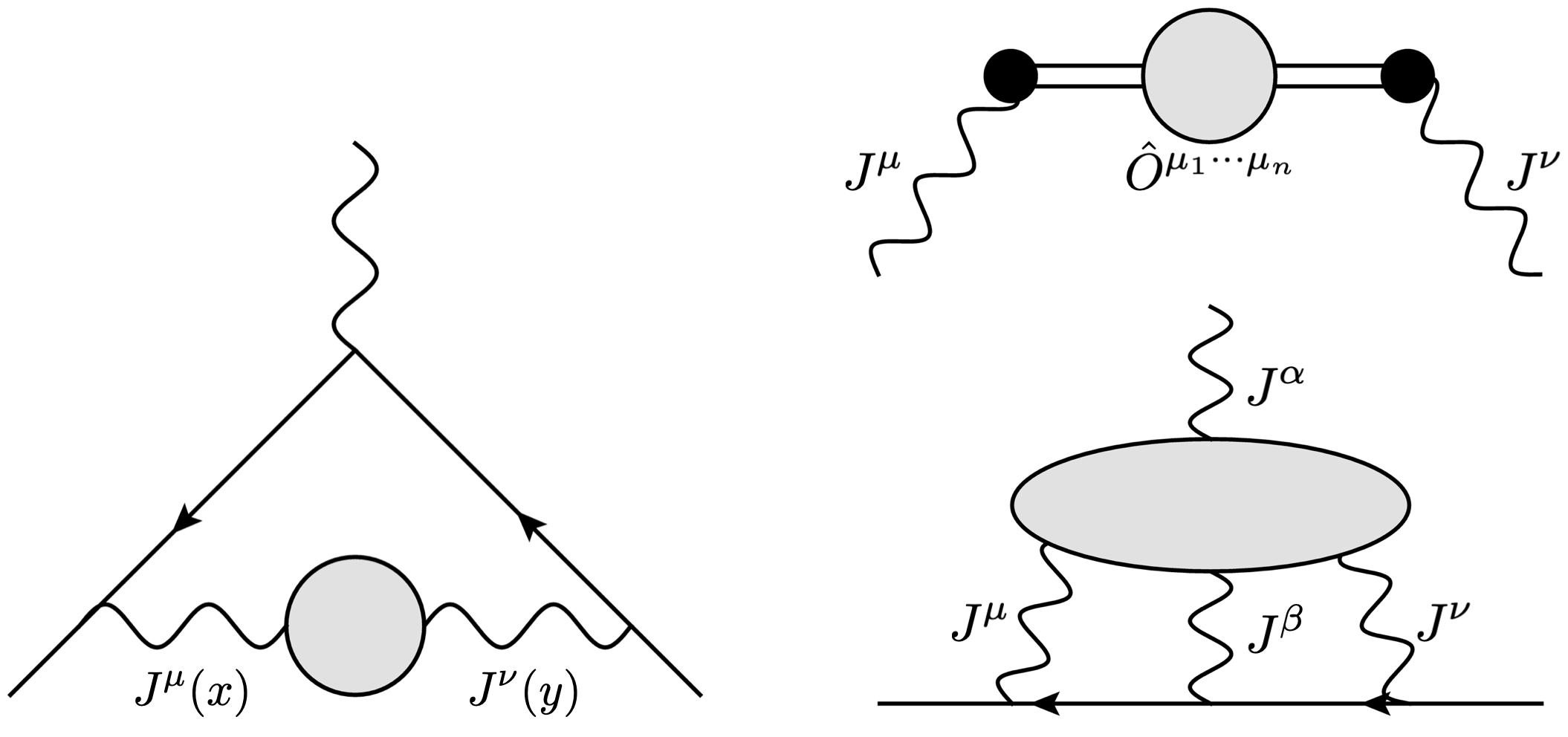}
    \caption{\textbf{Left:} HVP contribution to the muon's magnetic moment. \textbf{Right:} The HLBL contributions to the muon's magnetic moment (bottom), and the generic form of a matrix element contributing to the photon state (top). Electromagnetic currents are explicitly indicated, while the QCD elements are denoted by gray circles.}
    \label{fig:1}
\end{figure}

The dominant SM contribution to the anomalous moment comes from QED, and it has been computed to $\mathcal{O}(\alpha_{em}^5)$, with low theoretical uncertainty, where $\alpha_{em}$ is the QED coupling constant. Weak corrections have been evaluated up to two loops
but give a small numerical contribution since they are suppressed by powers of the muon to the $W$ boson mass ratio. By far the most complex and least controlled contributions originate from the QCD sector, where intrinsically non-perturbative effects come into play, see e.g.~\cite{Asmussen:2018oip,Jegerlehner:2017lbd}. At the leading order in $\alpha_{em}$, these contributions come from the hadronic vacuum polarization (HVP), associated to photons fluctuating into low energy hadronic states ($\mathcal{O}(\alpha_{em}^2)$, see Fig.~\ref{fig:1}), and hadronic light-by-light (HLBL) scattering, which involves photons combining into QCD loops annihilating to light ($\mathcal{O}(\alpha_{em}^3)$, see Fig.~\ref{fig:1}). 

State of the art theoretical calculations of the QCD contributions involve either the use of dispersive methods or euclidean lattice approaches~\cite{Jegerlehner:2017lbd}. Dispersive methods make use of the fact that a virtual photon decaying to hadrons can be related to $e^+e^-$ annihilation to QCD states, and that the hadronic spectral functions can be extracted from QCD decays. As a result, this is an intrinsically data driven approach. In contrast, in lattice based calculations one can directly extract correlators of electromagnetic currents, which are related to the HVP and HLBL. Traditional lattice QCD calculations are carried out in Euclidean time and, therefore, only limited HVP with spacelike momenta transfers. The spacelike lattice QCD results can be analytically continued to the timelike region only within the kinematic domains where no physical intermediate states are allowed. For the HVP this domain is limited by twice the pion mass, which vanishes in the chiral limit, for more detailed discussions see~\cite{Ji:2001nf,Ji:2001wha}. On the other hand, presently there are significant discrepancy between the results of hadronic contribution to HVP from lattice QCD calculations~\cite{Boccaletti:2024guq} and those estimated from the fits to the multiple experimental results over an extended timelike region~\cite{Aoyama:2020ynm}. Resolving this discrepancy is critical in determining whether BSM physics contribute to the muon's $g-2$. In light of this, it is very much welcome to have non-perturbative theoretical technique that can not only access HVP in an extended timelike region and at the same time capable of continuing to the spacelike region.

In this work we propose using quantum computers (QCs) to circumvent some of the issues found in Euclidean lattice calculations. The main motivation to using QCs is that they allow for direct computations in Minkowski time without the presence of a sign problem for any spacetime separation. In the current stage of QC hardware, however, the available quantum devices are still too small and error-prone to effectively perform any meaningful calculations in QCD, and any precision calculation, if possible, will only be achieved in some far future~\cite{Bauer:2022hpo,DiMeglio:2023nsa,Preskill:2018jim}. As a result, here we illustrate the approach in 1+1-d QED, i.e. the Schwinger model~\cite{Schwinger:1951nm}. This theory shares several phenomenological properties with 3+1-d QCD (see e.g.~\cite{Casher:1974vf,Coleman:1976uz,Banks:1975gq}), but it is technically simpler to implement on the lattice~\cite{Hamer:1997dx,Rothe:1992nt}. In particular, as we detail below, it can be efficiently analyzed using (classical) tensor network methods, see e.g.~\cite{Banuls:2019bmf}, as a precursor to implementations in large scale digital QCs~\cite{Farrell:2024fit,Farrell:2023fgd}.\footnote{See for example~\cite{Rigobello:2021fxw,Banuls:2024oxa,Papaefstathiou:2024zsu,Belyansky:2023rgh,Ikeda:2024rzv,Batini:2024zst,Silvi:2014pta} for further related studies of the gauge theories using quantum computing and tensor ntwork methods.} We recall that, despite the classical nature of tensor networks, their implementation mimics several aspects faced when using QCs, and thus both approaches are intimately connected, as we illustrate below.

Our calculations are performed in the massive Schwinger model; for the HVP the main object of our computation is the current-current correlator (see Fig.~\ref{fig:1})
\begin{align}\label{eq:hadronic_tensor}
    \Pi^{\mu \nu}(\vec x,\vec x_0)\equiv \langle T J^\mu(t,x)J^\nu(t_0,x_0)\rangle \, ,
\end{align}
where $t$ denotes time and $x$ the spatial direction, while $\vec x=(t,x)$. In QCD $J^\mu$ would correspond to the electromagnetic current for the appropriate combination of quark flavors;  in the single flavor  Schwinger model it is simply given by 
\begin{align}
 J^{\mu}(t,x) = \bar \psi(t,x) \gamma^\mu  \psi(t,x)   \, ,
\end{align}
where $\psi$ is a continuum two-component fermionic field, and $\mu=0,1$. The HVP tensor is then defined in momentum space as
\begin{align}\label{eq:main_PI}
 \Pi^{\mu\nu}(Q)&= \int d^2 \vec x \int d^2 \vec x_0 e^{i \vec Q \cdot (\vec x-\vec x_0)} \langle J^\mu(t,x)J^\nu(t_0,x_0)\rangle    \nn 
 &= (Q^\mu Q^\nu-\delta^{\mu\nu}Q^2)\Pi(Q^2)\, ,
\end{align}
where we used that the HVP only depends on $\vec x -\vec x_0$.\footnote{For further discussion on the hadronic tensor and related quantities in the quantum computing context see e.g.~\cite{Lamm:2019uyc,Mueller:2019qqj,Li:2021kcs,Qian:2021jxp}.} Similarly, the HLBL contribution is related to four point functions of the electromagnetic currents:
\begin{align}\label{eq:C4_def}
   \mathcal{C}^{\mu \alpha \beta \nu } =   \langle T J^\mu(\vec x) J^\alpha(\vec w) J^\beta(\vec y) J^\nu(\vec x_0) \rangle \, ,
\end{align}
corresponding to the simplest realization of Fig.~\ref{fig:1}, when $\hat Q^{\alpha \beta}=J^\alpha J^\beta$.

In this paper, we evaluate Eqs.~\eqref{eq:main_PI} and~\eqref{eq:C4_def} from real-time lattice simulations. As alluded before, the computations are performed using (classical) tensor network methods, which do not suffer from sign problems related to real-time evolution. More, they offer a first step towards the implementation in quantum computers, where larger scale and higher dimensional quantum simulations should eventually become feasible in the near future. 
To that end, we first review some basic properties of the Schwinger model in section~\ref{sec:Schwinger_model}, and then present numerical results from tensor network simulations for the extraction of the HVP and HLBL tensors in sections~\ref{sec:HVP},~\ref{sec:HLBL}. We further illustrate our approach by providing a calculation using a QC emulator in a small computational lattice. We conclude in section~\ref{sec:conclusions} by summarizing our results and discussing future extensions of this work.

\section{The Schwinger model}\label{sec:Schwinger_model}
The Schwinger model corresponds to the theory of quantum electrodynamics  in 1+1-d~\cite{Schwinger:1951nm}. In the temporal gauge, i.e. $A^0=0$, the Hamiltonian of this model is given by
\begin{align}\label{eq:H_Schwinger}
	H&=  \int dx \, \frac{1}{2 } E^2(x) \nn 
 &+ \bar  \psi(x) (-i \gamma^1 \partial_1 +g \gamma^1 A_1(x) +m ) \psi(x)\, ,
\end{align}
where $\psi$ denotes the single flavor two component spinor field with mass $m$, while $A^\mu$ is the gauge field, of which only $A^1$ appears explicitly in $H$. The electric field, $E=-\partial_t A^1$, carries all the energy associated to the gauge degrees of freedom. Note that in two dimensions there is no magnetic field, and, in fact, the gauge field is not dynamical and acts as a potential. To make explicit the static character of the gauge field, one can integrate out the remaining field component using Gauss's law: $   \partial_x E=g  \psi^\dagger \psi$.

The theory can be discretized on a lattice, following the Kogut-Susskind prescription for fermions~\cite{Susskind:1977lf,Kogut:1974ag}. In practice, this is done by introducing the one component spinor $\chi_n$, which is related to the continuum one by 
\begin{align}\label{eq:help_1}
   \psi(\tilde n )  \to \frac{1}{\sqrt{a}}\begin{pmatrix}
		\chi_{2 \tilde n}\\
		\chi_{2 \tilde  n-1}
	\end{pmatrix}\, .
\end{align}
with $a$ the staggered lattice spacing.\footnote{We denote the naive lattice spacing $\tilde a=2 a$.} Here we use $\tilde n$ as the index on the physical lattice; we shall use $n$ as the site index on the staggered lattice. For each $\tilde n$ there are two values of $n$. The gauge field is introduced in terms of the reduced electric field $ L\equiv  E/g$ and can be integrated out by using the discrete Gauss' law generator
 \begin{align}
G_n&=\delta L_n-  \Bigg( \chi_n^\dagger \chi_n -\frac{1}{2}(1-(-1)^n)\Bigg)\, ,
\end{align}
where $\delta L_n\equiv  L_n-L_{n-1}$. In the charge zero sector, one has that any physical state satisfies $G|\psi\rangle =0$.

The single component spinors can be mapped to spin operators via a Jordan-Wigner transform (JWt)
\begin{align}
	&\chi_n =   S_n \sigma^-_n\, , \quad  \chi^\dagger_n =   S^\dagger_n \sigma^+_n\, ;
\end{align}
where the string operator reads $ S_n\equiv   \prod_{ k< n}[(-i)\sigma^z_k ]$, and $\sigma^{x,y,z,+,-}_n$ denote the standard Pauli matrices acting on the site $n$. We define $\sigma^{\pm} = (\sigma^x\pm i \sigma^y)/2$. Using the lattice discretization described and employing the JWt, one finds that the continuum Hamiltonian in Eq.~\eqref{eq:H_Schwinger} maps to the spin Hamiltonian
\begin{align}\label{eq:Spin_Hamiltonian}
	H &= 	\frac{g^2a}{2} \sum_{n=1}^{N-1} \left[\frac{1}{2}\sum_{k=1}^n (\sigma^z_k+(-1)^k) \right]^2  +\sum_{n=1}^{N} m  (-1)^n \frac{\sigma^z_n}{2} \nn
	&+\frac{1}{2a} \sum_{n=1}^{N-1}  \sigma^+_n \sigma^-_{n+1} + \sigma^+_{n+1}  \sigma^-_n  \, ,
\end{align}
where we imposed open boundary conditions, such that for the $N$th lattice point $L_N=\chi_{N+1}=0$. For the $\gamma$ matrices we used the basis $\gamma^0=\sigma^z$, $\gamma^1 = -i \sigma^y$, and $\gamma^5=-\sigma^x$.

Finally, having provided the lattice version of the Schwinger model, we turn our attention to the form of the current operators on the lattice. In what follows, we shall only need the temporal component of the electric charge density, $J^0$, which reads
\begin{align}
  \psi^\dagger(\tilde n) \psi( \tilde n )    &= \frac{1}{a} \left( \chi_{2 \tilde n}^\dagger \chi_{2 \tilde n} +\chi_{2 \tilde n-1}^\dagger \chi_{2 \tilde n-1} \right) \nn 
  &=  \frac{1}{a} (\sigma^+_{2 \tilde n} \sigma^-_{2\tilde n}+ \sigma^+_{2 \tilde n-1} \sigma^-_{2 \tilde n-1} )   \, .
\end{align}

\section{Hadronic vacuum polarization}\label{sec:HVP}
Having introduced the Schwinger model, we now discuss how to extract the HVP in this theory. In general, one can rewrite the position space form of the current-current correlator as
\begin{align}
  \Pi^{\mu \nu} = \langle J^\mu(t,x)  J^\nu (0,x_0)\rangle  \equiv \langle \psi^\mu_l|\psi^\nu_r \rangle\, ,
\end{align}
with
\begin{align}\label{eq:L_and_R_states}
   |\psi_l^\mu \rangle &=    j^\mu(\tilde n) e^{-iHt} |\Omega\rangle \, ,\nn 
   |\psi_r^\nu\rangle &=  e^{-iHt}  j^\nu(\tilde n_0) |\Omega\rangle \, ,
\end{align}
where $j^\mu$ denotes the discretized version of the current operator and $|\Omega \rangle$ is the ground state of the Schwinger model at finite $g$ and $m$. To perform the Fourier transform necessary to obtain Eq.~\eqref{eq:main_PI}, we use
\begin{align}\label{eq:HVP_lattice}
 &\Pi^{\mu \nu}(Q^0,Q^1)=  \int d x \,dt \, e^{i Q^0 t} e^{-i Q^1 x} \Pi^{\mu \nu}(t,x) \nn 
 &= \sum_{j=0}^{N_x-1} \sum_{k=0}^{N_t-1} \, \Delta t \,  \Delta x \, e^{i Q^0 k \Delta t} e^{-i Q^1 j \Delta  x} \Pi^{\mu \nu}(k \Delta t,j \Delta x  ) \, ,
\end{align}
where $N_t$ and $N_x$ denote the number of points taken in the time and spatial directions, respectively. Here $\Delta t$ and $\Delta x$ correspond to the spacing in the time and spatial directions; the respective momentum modes read $Q^0 = 0 , \frac{2\pi}{N_t \Delta t  }  , \cdots \frac{2\pi}{N_t \Delta t  } (N_t-1)$ and $ Q^1 = 0 , \frac{2\pi}{N_x \Delta x  }  , \cdots \frac{2\pi}{N_x \Delta x  } (N_x-1)$.

We start by discussing the form of Eq.~\eqref{eq:HVP_lattice} in the exactly solvable limits of the Schwinger model. The first limit corresponds to the case where $m=0$ and the theory can be shown to become that of free massive bosons, with a mass $m_{b}= g/\sqrt{\pi}$~\cite{Coleman:1976uz}. In this case one has that~\cite{Blaschke:2014ioa}
\begin{align}
\Pi(Q^2)\Big\vert_{m=0} = \frac{1}{\pi \left(Q^2+m_b^2\right)} \, .
\end{align}
The HVP is regular for spacelike $Q$, but has a pole at timelike separations at the value of the boson mass. Another interesting limit corresponds to the case where one decouples the fermion sector from the gauge field, which can be achieved in the $g\to 0$ limit, or by treating $A^\mu$ as an external field. In this case, one has that~\cite{Blaschke:2014ioa}
\begin{align}
\Pi(Q^2)\Big\vert_{g=0} &=  \frac{1}{\pi Q^2} +  \frac{4 m^2 \arctan\left(\frac{\sqrt{Q^2}}{\sqrt{-4 m^2-Q^2}}\right)}{\pi  \sqrt{Q^{6}} \sqrt{-4 m^2-Q^2}} \nn 
&\stackrel{m\to 0}{\to} \frac{1}{\pi Q^2} \, . 
\end{align}
In this case there is a pole associated to the massless limit of the theory, while when $m\neq 0$, there is a branch cut along the timelike axis related to the particle production threshold.

Finally, to go beyond these two cases, when $m$ and $g$ are taken finite, one has to extract $\Pi^{\mu \nu}$ from the lattice, which we follow to do using tensor network methods. In the what follows we use $a\cdot g=0.50$, $a\cdot m=0.25$, and we consider the lattice improved action by shifting the mass term as $  m_{\rm lat} = m_{\rm bare} -g^2 a/8 $~\cite{Dempsey:2022nys}. We will focus on $\Pi^{00}$; the extraction of the two other components poses no extra difficulties. In particular, note that $ \Pi(Q^2) =  \Pi^{00}(Q^2)/Q_1^2$ for $Q$ not timelike. 

\begin{figure}[h!]
    \centering
    \includegraphics[width=0.45\textwidth]{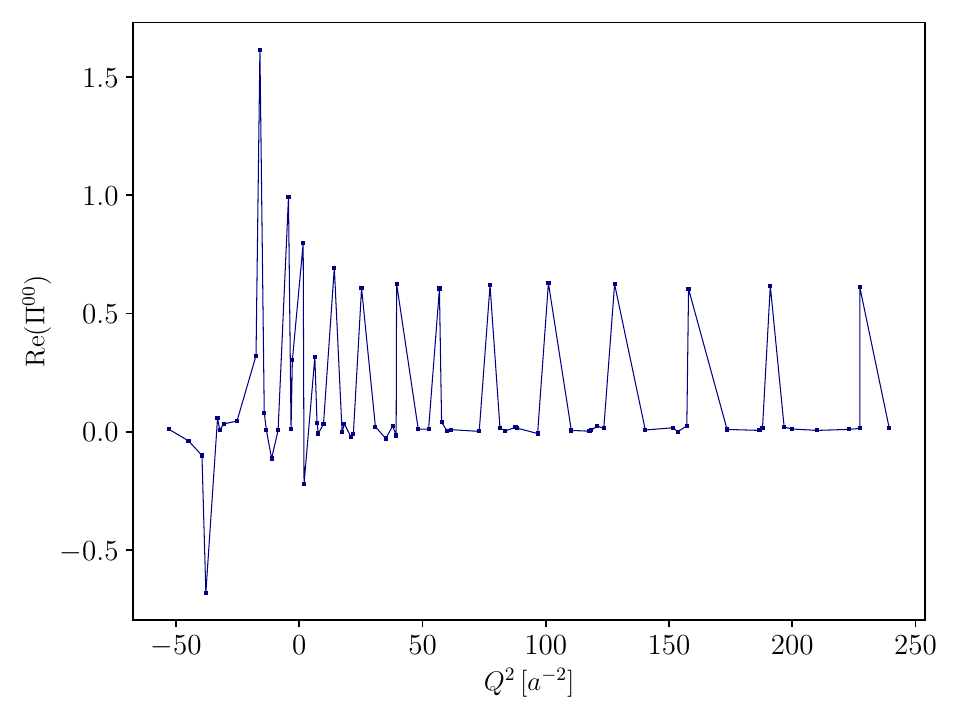}
    \caption{HVP extracted from the tensor network simulations using MPS. The quantities are plotted in terms of the lattice spacing $a$.}
    \label{fig:DMRG_result}
\end{figure}

In Fig.~\ref{fig:DMRG_result} we show the HVP extracted from a tensor network simulation. We used the tensor network package \texttt{ITensor}~\cite{itensor} in \texttt{Julia}, adopting a matrix product state (MPS) form to represent the quantum states. The time evolution of the states in Eq.~\eqref{eq:L_and_R_states} is achieved using the time-dependent variational principle (TDVP) algorithm~\cite{Haegeman:2011zz,PhysRevB.94.165116}, while the preparation of the initial state of the system is performed using the density matrix renormalization group (DMRG) algorithm~\cite{PhysRevLett.69.2863,PhysRevB.48.10345}. We utilized a lattice with $N=120$ staggered sites, and we extract the observables in the center of the lattice, minimizing edge effects, which are suppressed by the finite fermion mass. We use six values of spatial separation for $ x -  x_0 = 0,1,\cdots 5$ in units of the lattice spacing, and evolve the states up 
sufficiently large times, $0<t-t_0<5$ in lattice spacing units, reducing the time steps to ensure that the trotterization of the evolution operator does not significantly affect the final distribution. The result shown exhibits the expected feature for the parameters used, i.e. a decay in both the spacelike and timelike directions, with a peak around the origin. Nonetheless, we note that the obtained curve's quality depends greatly on the numerical Fourier transform one has to perform. Nonetheless, this operation is purely classical, and even when using a QC, it would be realized the same way as illustrated here. Thus, the only requirement to obtain a better numerical result for $\Pi^{00}$, lies on increasing (decreasing) the lattice size (spacing) and collecting more data points. 

\begin{figure}[H]
    \centering
    \includegraphics[width=0.45\textwidth]{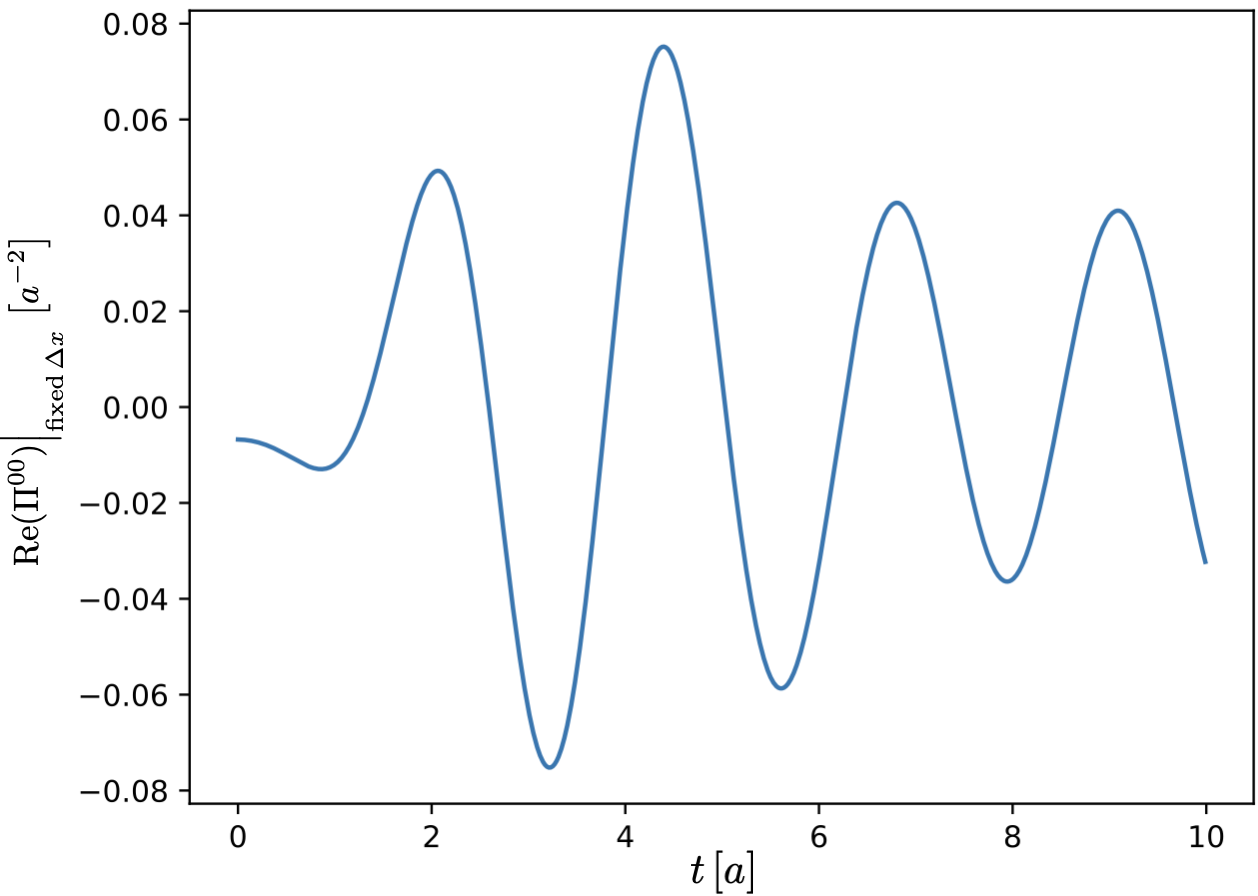}
    \caption{Time evolution of $\text{Re}(\Pi^{00})$, extracted from classical emulator of a digital quantum computing, for $\Delta x= x -  x_0 =2$ in lattice units; the remaining parameters are the same as the ones used in Fig.~\ref{fig:DMRG_result}.}
    \label{fig:current-current}
\end{figure}

In Fig.~\ref{fig:current-current} we show the real-time quantum simulation of $\Pi^{00}$ (in position space) for $\Delta x=x -  x_0 =2$ in lattice units, in a smaller system with $N=20$ qubits. Unlike the above classical real time simulations, where the ground state can be directly obtained using the DMRG algorithm, constructing the ground for a quantum simulation is, in general, at least as hard as performing the subsequent real time evolution. In the present simulations we circumvent this issue by performing the simulation using the built-in classical emulator of a quantum computer in \texttt{Qiskit}. This allows the ground state to be directly converted into a quantum circuit, while the state itself can be directly obtained for the small system we consider.\footnote{By this we mean that in a small system, the current approach is equivalent to directly diagonalizing the Hamiltonian, and thus we have access to the complete Hilbert space.} Alternatively, we could have, for this small system, loaded the state from a DMRG run using tensor networks, or more generally used a hybrid variational method, such as VQE, to obtain this state.

\begin{figure}[H]
    \centering
    \includegraphics[width=\linewidth]{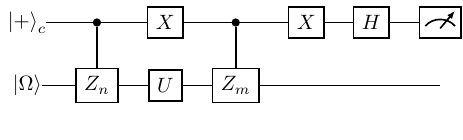}
    \caption{Measurement protocol used to extract $\Pi^{00}$. Here, $X$ and $H$ operators denote the $\sigma^x$ and Hadamard gates, respectively.}
    \label{fig:meas}
\end{figure}

We again recall that the transformation to momentum space can be done the same way as shown for the above MPS example. However, one important difference compared to tensor network methods relates to the extraction of the expectation value of hermitian operators, since quantum computers only \textit{naturally} allow for the application of unitary gates (operators). In the example shown in Fig.~\ref{fig:current-current} this is done by using the simple interference measurement protocol, i.e. the Hadamard test, illustrated in Fig.~\ref{fig:meas}. Taking the initial state of a quantum simulation, $ | \Omega\rangle$, and adding a single ancilla qubit in the $|+\rangle=(|0\rangle+|1\rangle)/\sqrt{2}$, one can apply a series of gates to generate a characteristic interfering pattern:
\begin{align}\label{eq:help1}
 |\Omega\rangle  |0\rangle_c  \to \frac{Z_mU}{\sqrt{2}}\ket{\Omega}\ket{0}_c+\frac{UZ_n}{\sqrt{2}}\ket{\Omega}\ket{1}_c \, ,
\end{align}
where $U=e^{-itH}$ is the time evolution operator and $Z_m$ denotes the relevant operator in the lattice discretiz<tion of the current operators. The evolution in Eq.~\eqref{eq:help_1} is implemented by the first four gates in Fig.~\ref{fig:meas}. The last two gates in the circuit diagram implement the measurement of the ancilla in the $X$-basis, which allow to directly compute $\text{Re}\bra{\Omega}Z_m(t)Z_n\ket{\Omega}$. A \texttt{Qiskit} code to compute this function is provided in \cite{code}, along with more details.

\section{HLBL and hadronic structure of the photon}\label{sec:HLBL}
Having computed the HVP, we now consider the HLBL contribution. This involves the (Fourier transform of) expectation values of the form shown in Eq.~\eqref{eq:C4_def}, see also Fig.~\ref{fig:1}, which can be extracted in a similar fashion to the HVP, despite the more complex numerical calculation. We note that the HLBL contribution is closely related to the correlators determining the photon's hadronic structure~\cite{Jager:2005uf,Ji:2001nf,Ji:2001wha}. In this case, and in QCD, one is interested in computing the expectation value of a quark/gluon
local operator, in between virtual photon states, see Fig~\ref{fig:1}. For example, at leading order in $\alpha_{em}$ and in 3+1-d, this results in
\begin{align}\label{eq:hhh}
   \mathcal{C}^{\mu \mu_1\cdots \mu_n \nu }(\vec x - \vec x_0)   \varepsilon^\mu(\lambda)\varepsilon^{*,\mu}(\lambda) \, ,
\end{align}
where $\mathcal{C}^{\mu \mu_1\cdots \mu_n \nu } = \langle J^\mu(\vec x) \hat O^{\mu_1 \cdots \mu_n}(0) J^\nu(\vec x_0) \rangle $ is illustrated in Fig.~\ref{fig:1}, with $\hat O$ denoting the local operator being measured, and 
$\varepsilon^\mu$ is the photon polarization vector, which has $d-1$ physical polarizations. For $d=1$ there are no physical polarizations (the electric field is not propagating), and the contraction in Eq.~\eqref{eq:hhh} vanishes exactly. Nevertheless, it is still possible to compute the correlator $\mathcal{C}$; to that end we consider the simplest case where the local operator is given by the product of two electromagnetic currents:
\begin{align}\label{eq:HLBL_final}
   \mathcal{C}^{\mu \mu_1 \mu_2 \nu }_4(\vec x - \vec x_0)  = \langle  T J^\mu(\vec x) J^{\mu_1}(0)J^{\mu_2}(0)J^{\nu}(\vec x_0)\rangle \, ,
\end{align}
corresponding to Eq.~\eqref{eq:C4_def}, with $\vec w=\vec y=0$.

\begin{figure}[h!]
    \centering
    \includegraphics[width=0.45\textwidth]{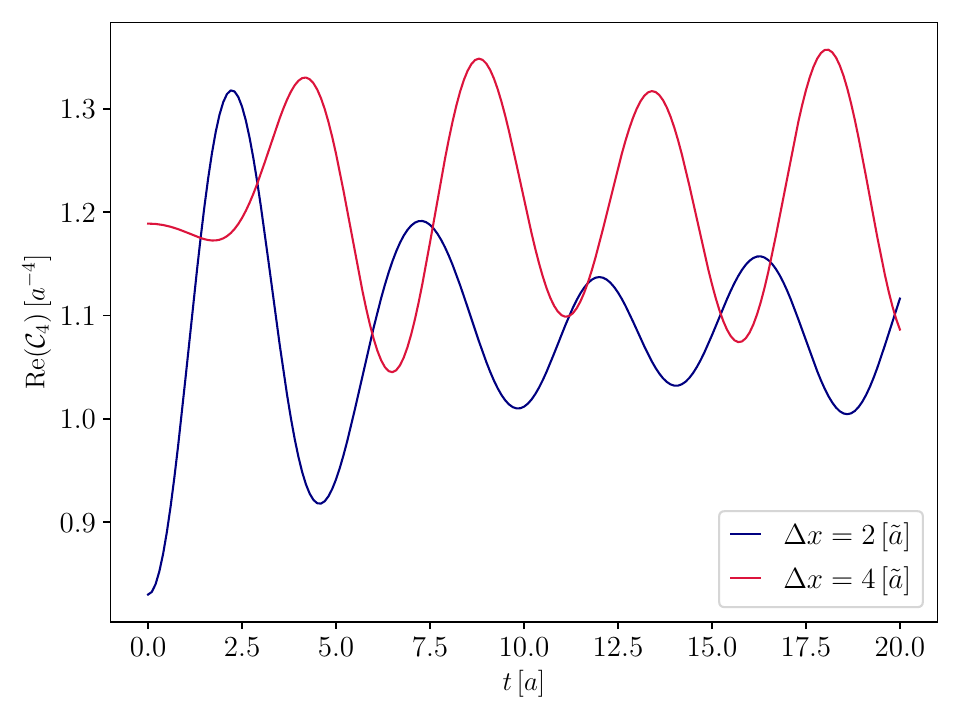}
    \caption{Four point current correlator extracted from a tensor network simulation, following Eq.~\eqref{eq:HLBL_final} with $\mu=\nu=\mu_1=\mu_2=0$. Here $\Delta x$ denotes the spatial separation, while the horizontal axis denotes the time evolution of the expectation value.}
    \label{fig:4pt_result}
\end{figure}

In Fig.~\ref{fig:4pt_result} we show the evaluation of Eq.~\eqref{eq:HLBL_final} using a MPS tensor network, for $\Delta x= x-x_0=2,4$ in units of $\tilde a$ as a function of time ($t_0=0$). We use a lattice with $N=60$ sites in a set up analog to the HVP case, and the intermediate current operators are evaluated in the middle of the lattice. Numerically, we observe that the extraction of this expectation value is considerably more complex than the HVP case, with the bond dimensions of the intermediate links in the MPS growing much faster with time. This limits the amount of data that can be extracted efficiently in our simple evaluation. We, therefore, refrain from performing the final Fourier transform to momentum space.

\section{Conclusion and discussion}\label{sec:conclusions}

In this work we have explored the potential use of quantum computing methods and technologies to extract the HVP from timelike momenta transfers. These calculations can be extended to access the HBLB as well as the hadronic structure of the photon. While these computations are related to the structure of hadrons~\cite{Lamm:2019uyc,Grieninger:2024cdl,Grieninger:2024axp}, we note an important difference between studying the photon structure and the partonic distributions inside hadrons: the latter will require preparing a hadronic target state, which is usually hard to obtain through quantum computing methods; the former has as external states virtual photons, which can be easily created from the vacuum by the insertion of electromagnetic currents.

In order to illustrate the main points of our discussion, we have provided a small system size tensor network calculation of the HVP and HLBL in the Schwinger model. The major difficulty in the extraction of these quantities lies in the need to Fourier transform to momentum space, requiring a finer spatial and time grid than the ones employed here. We hope to address these numerical issues in the future, with more precise numerical calculation. We further note that when translating these computations to a real quantum computer, one may face further difficulties, such as the extraction of the expectation value of (hermitian) observables~\cite{Lamm:2019uyc}, which are not present using tensor networks; this is one of the important differences between these two approaches. We further illustrated this point, by extracting the position space HVP using a quantum simulator.

The methods employed here surpass some of the issues faced using either dispersive relations or euclidean lattice methods, but they require a significant progress in a formulation of QCD which could be implemented in real quantum computers. We hope coming developments in the field will allow for the questions and methods posed here to be tested in a context closer to QCD or in other field theories, see e.g.~\cite{ Dempsey:2023fvm,Czajka:2022plx,PhysRevD.109.016023,Barata:2023jgd,Klco:2018kyo, Butt:2019uul, Magnifico:2019kyj, Kharzeev:2020kgc,PhysRevLett.132.091903,Pedersen:2023asd}.

\begin{acknowledgments}
We thank A. Florio, R. Pisarski, A. V. Sadofyev, and R. Szafron for helpful discussions related to this work. This material is based upon work supported by the U.S. Department of Energy, Office of Science, Office of Nuclear Physics and National Quantum Information Science Research Centers, Co-design Center for Quantum Advantage (C2QA) under contract number DE-SC0012704. The work of S.M. is partially supported by Laboratory Directed Research and Development (LDRD) funds from Brookhaven Science Associates. The work of J.R. was supported by the U.S. Department of Energy, Office of Science, Office of Workforce Development for Teachers and Scientists (WDTS) under the Science Undergraduate Laboratory Internships Program (SULI) and the Brookhaven National Laboratory (BNL), Physics Dept. under the BNL Supplemental Undergraduate Research Program (SURP).

\end{acknowledgments}

%%%%%%%%%%%%%%%%%%%
\bibliographystyle{elsarticle-num}

\bibliography{references.bib}

\end{document}